\documentclass[twocolumn,showpacs,preprintnumbers,amsmath,amssymb,superscriptaddress]{revtex4-1}

\usepackage{graphicx}
\usepackage{dcolumn}
\usepackage{bm}
\usepackage{color}
\usepackage{ulem}
\usepackage{pifont}
\usepackage{natbib}
\usepackage{hyperref}
\hypersetup{
colorlinks = true,
urlcolor   = blue,
linkcolor  = blue,
citecolor  = blue
}

\begin{document}

\preprint{APS/PRB}
\title{Co-existence of short- and long-range magnetic order in LaCo$_2$P$_2$}

\author{Ola Kenji Forslund}
    \email{okfo@kth.se}
\affiliation{Department of Applied Physics, KTH Royal Institute of Technology, SE-106 91 Stockholm, Sweden}
\author{Daniel~Andreica}
\affiliation{Ioan Ursu Institute, Faculty of Physics, Babes-Bolyai University, 400084 Cluj-Napoca, Romania}
\author{Hiroto~Ohta}
\affiliation{Faculty of Science and Engineering, Doshisha University, Kyotanabe, Kyoto 610-0321, Japan}
\author{Masaki~Imai}
\author{Chishiro~Michioka}
\author{Kazuyoshi~Yoshimura}
\affiliation{Department of Chemistry, Graduate School of Science, Kyoto University, Kyoto 606-8502 Japan}
\author{Martin M\aa nsson}
    \email{condmat@kth.se}
\affiliation{Department of Applied Physics, KTH Royal Institute of Technology, SE-106 91 Stockholm, Sweden}
\author{Jun Sugiyama}
\affiliation{Neutron Science and Technology Center, Comprehensive Research Organization for Science and Society (CROSS), Tokai, Ibaraki 319-1106, Japan}
\affiliation{Advanced Science Research Center, Japan Atomic Energy Agency, Tokai, Ibaraki 319-1195, Japan}
\date{\today}

\begin{abstract}
The ferromagnetic (FM) nature of the metallic LaCo$_2$P$_2$ was investigated with the positive muon spin rotation, relaxation and resonance ($\mu^+$SR) technique. Transverse and zero field $\mu^+$SR measurements revealed that the compound enters a long range FM ground state at $T_{\rm C}=130.91(65)$~K, consistent with previous studies. Based on the reported FM structure, the internal magnetic field was computed at the muon sites, which were predicted with first principles calculations. The computed result agree well with the experimental data. Moreover, although LaCo$_2$P$_2$ is a paramagnet at higher temperatures $T>160$~K, it enters a short range ordered (SRO) magnetic phase for $T_{\rm C}<T\leq160$~K. Measurements below the vicinity of $T_{\rm C}$ revealed that the SRO phase co-exists with the long range FM order at temperatures $124\leq T\leq T_{\rm C}$. Such co-existence is an intrinsic property and stems from competition between the 2D and 3D interactions/fluctuations.

\end{abstract}


\keywords{short keywords that describes your article}

\maketitle

\section{\label{sec:Intro}Introduction}
The interplay between magnetism and superconductivity is a long-term popular problem, particularly since the discovery of high-$T_c$ cuprates, because the ground states are usually considered incompatible with each other \cite{Bardeen1957}. While the coexistence of both phases have been reported \cite{Lange2003, Marsik2010}, a competition between the two phases is more common and has been observed in many systems such as rare earth ($R$) $R$Ni$_2$B$_2$C \cite{Eisaki1994}, K doped Ba$_{1-x}$K$_x$Fe$_2$As$_2$ \cite{Rotter2009} or BaFe$_{1.89}$Co$_{0.11}$As$_2$~\cite{Marsik2010}. The latter two compounds in particular crystallize in a ThCr$_2$Si$_2$-type structure, for which the general structure is described by $AT_2X_2$ with a metal $A$, transition metal $T$ and metalloid $X$ atoms. In these systems, the interactions are presumably dominated by low dimensional fluctuations within the edge-sharing $TX_4$ tetrahedra layers. The interlayer interaction across the $A$ layer is heavily dependent on the $X-X$ bonding distance, which naturally changes depending on the specific elements that occupy each site of $AT_2X_2$. As a result, many ground states, such as, paramagnetism (PM), ferromagnetism (FM), antiferromagnetism (AF), short range order and superconductivity have been reported for $AT_2X_2$ \cite{Reehuis1990, Baumbach2014, Thompson2014, Sugiyama2015}, 
since this family can accommodate many elements and combinations. The flexibility in the combination of constituent elements results in ground states that greatly varies with chemical doping~\cite{Thompson2014, Sugiyama2015, Xiaoyan2016} and by application of hydrostatic pressure \cite{Baumbach2014}. 

\begin{figure}[ht]
  \begin{center}
    \includegraphics[keepaspectratio=true,width=85 mm]{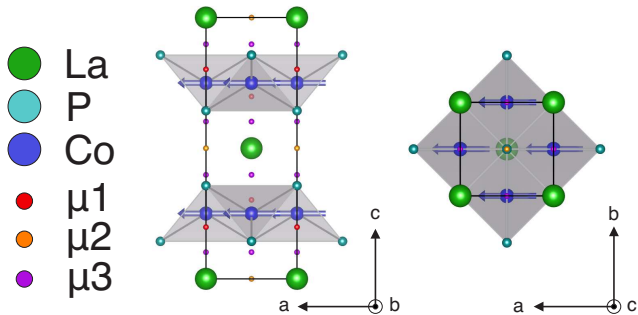}
  \end{center}
  \caption{The magnetic structure of LaCo$_2$P$_2$ proposed by Ref.~\onlinecite{Reehuis1994}. The predicted muon sites by DFT calculations are indicated as spheres: red $\mu1$ at (0,0,0.198), orange $\mu2$ at (0,0.5,0.102) and purple $\mu3$ at (0,0,0.5). The coordinates are specified for the crystal space group  $I4/mmm$ ($\#$139): $a=3.812$, $b=3.812$, $c=10.984~$\AA. The magnetic moments are indicated as blue arrows within the CoP-polyhedra.  
  }
  \label{fig:La_Crystal}
\end{figure}

It has been observed in several $AT_{2}$P$_{2}$ ($A$ = Ca, Sr, and Ba, and T = Fe, Co, and Ni) compounds that the magnetic phase transitions are related to subtle structural changes in the crystals \cite{Hoffman1985, Reehuis1990, Jia2009}. These compounds have shown to transform from an uncollapsed tetragonal (ucT) phase to a collapsed tetragonal (cT) phase as a function of chemical doping \cite{Reehuis1990, Jia2009, Sugiyama2015}, which naturally affects the ground state. 
While the majority of the $AT_{2}$P$_{2}$ compounds exhibit AF or PM ground states, LaCo$_2$P$_2$ is an itinerant FM with $T_{\rm C}=125$~K. Early magnetization measurements showed a Curie-Weiss behavior at higher temperatures above $T_{\rm C}$ \cite{Morsen1988, Reehuis1990}. Neutron diffraction measurements on a polycrystalline sample revealed a collinear FM structure along the $a$-axis with an ordered moment of $0.44(3)~\mu_{\rm B}$ (Fig.~\ref{fig:La_Crystal}) \cite{Reehuis1994}. Indeed, a recent magnetization measurements, performed on a single crystal at low temperature, suggests a highly anisotropic FM ground state \cite{Imai2015}. Meanwhile, the spin-lattice relaxation rate reported in a $^{31}$P-NMR study \cite{Imai2015} indicated a 3D character of the spin fluctuations above $T_{\rm C}$. 


Since LaCo$_2$P$_2$ exhibits a 3D character above $T_{\rm C}$, but a highly anisotropic character below $T_{\rm C}$, 
it is of high interest to study the crossover between these two regimes across $T_{\rm C}$. Therefore, we have initiated a detailed of LaCo$_2$P$_2$ using muon spin rotation, relaxation and resonance ($\mu^+$SR), taking advantage of the fact that $\mu^+$SR can offer information about the internal magnetic field distributions of both static and dynamic characters \cite{Hayano1979, Forslund2019, Sugiyama2021}. In this study, a cascade of magnetic transitions is observed using $\mu^+$SR. In particular, both short and long range magnetic ordered phases are likely to coexist in LaCo$_2$P$_2$ in the vicinity of $T_{\rm C}$. Such behavior is attributed to an intrinsic property and is most likely originating from the competing interactions of different dimensionality.

\section{\label{sec:exp}Experimental Setup}
Polycrystalline LaCo$_2$P$_2$ was prepared from pure La, and Co and P starting materials. LaP and Co$_2$P were first synthesized by a solid state reaction between La/Co and P in evacuated quartz tubes at 800$^\circ$C/700$^\circ$C. LaCo$_2$P$_2$ could then be synthesized from a solid state reaction between LaP and Co$_2$P, kept at 1000$^\circ$C for 20 hours in an Ar atmosphere. Details about the sample synthesis can be found elsewhere \cite{Sugiyama2015_1, Imai2014}.

The $\mu^+$SR measurements were performed at the Dolly instrument at the S$\mu$S muon source at Paul Scherrer Institute, Switzerland. A top loaded $^4$He cryostat was used in order to reach temperatures down to $\sim2$~K. 
About 500~mg of powder sample was inserted into an Al-coated Mylar tape envelope. The envelope was attached to a low-background fork type sample holder made of Cu. The  $\mu^+$SR data was analyzed using MUSRFIT \cite{musrfit}.

The muon sites and the local spin density at the muon sites in LaCo$_2$P$_2$ were predicted by density functional theory (DFT) calculations using a full-potential linearlized augmented plane-wave method within generalized gradient approximation as implemented in WIEN2k program package \cite{WIEN2k}. In the calculations, the lattice parameters and atomic positions of LaCo$_2$P$_2$ were taken from Ref.~\onlinecite{Reehuis1990}. The magnetic moments were aligned parallel to the [100] direction through spin-orbit coupling. The muffin tin potential radii ($R_\mathrm{MT}$) for La, Co, and P were taken to be 2.50, 2.36, and 1.84~\AA, respectively. The energy cutoff was chosen to be ($R_\mathrm{MT} \times K_\mathrm{max} = 7.0$, and 20 $\times$ 20 $\times$ 20 $k$-points meshes were used in the Brillouin zone. Here, $ K_\mathrm{max}$ is the maximum modulus for the reciprocal vectors.

\section{\label{sec:results}Results}
The presentation of the $\mu^+$SR results of LaCo$_2$P$_2$ is divided into sections based on the type of field configuration chosen for the $\mu^+$SR experiments: transverse field (TF) or zero field (ZF). Transverse refers to the applied field direction with respect to the initial muon spin polarization. Additionally, the ZF time spectra collected at base temperature is reproduced based on the published magnetic structure and muon site determined from DFT calculations.

\subsection{\label{sec:results}Transverse field}
\begin{figure}[ht]
  \begin{center}
     \includegraphics[keepaspectratio=true,width=65 mm]{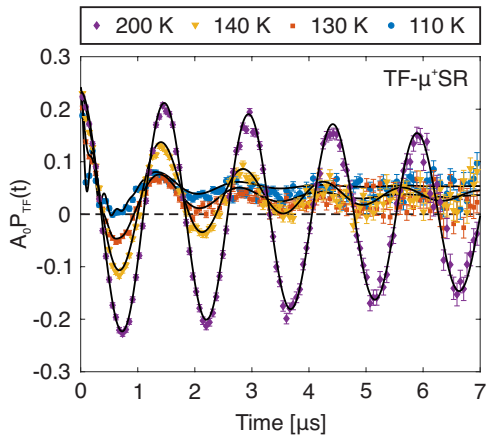}
 \end{center}
    \caption{
  Transverse field $\mu^+$SR spectra for selected temperatures 
  ($T=110$, 130, 140 and 200~K) for the LaCo$_{2}$P$_{2}$ compound. 
  Solid lines represent the best fit using Eq.~(\ref{eq:TF}).
  }
    \label{fig:La_TFSpec}
\end{figure}

Figure~\ref{fig:La_TFSpec} shows the collected TF ($\sim50$~Oe) spectra of LaCo$_{2}$P$_{2}$ for selected temperatures. At high temperatures, a distinct oscillation is observed with a frequency of about 0.7~MHz, corresponding to the applied field TF. As the temperature is lowered, the amplitude of the 0.7~MHz  oscillation, $i.e.$ the asymmetry due to the applied TF, decreases. Although, the amplitude is not completely diminished at lower temperature. Moreover, a faster oscillation can be observed in the time spectra at lower temperatures, accompanied by a positive offset. Therefore, the TF spectra were fitted using a combination of three exponentially relaxing cosine oscillations together with an non-oscillating exponential relaxation:

\begin{eqnarray}
 A_0 \, P_{\rm TF}(t) &=&
A^{\rm PM}_{\rm TF} \cos(f^{\rm PM}_{\rm TF}2\pi t+\phi^{\rm PM}_{\rm TF})e^{-\lambda^{\rm PM}_{\rm TF}t} \cr
&+& A^{\rm imp}_{\rm TF} \cos(f^{\rm imp}_{\rm TF}2\pi t+\phi^{\rm imp}_{\rm TF})e^{-\lambda^{\rm imp}_{\rm TF}t} \cr 
&+& A_{\rm FM} \cos(f_{\rm FM}2\pi t+\phi_{\rm FM})e^{-\lambda_{\rm FM}t} \cr
&+& A_{\rm S}e^{-\lambda_{\rm S} t},
\label{eq:TF}
\end{eqnarray}
where $A_{0}$ is the initial asymmetry determined by the the detector geometry of the instrument and $P_{\rm TF}$ is the muon spin polarization function in TF configuration. $A_{\rm TF}$, $f_{\rm TF}$, $\phi_{\rm TF}$ and $\lambda_{\rm TF}$ are the asymmetry, frequency, initial phase and depolarization rate resulting from the applied TF, where the superscripts PM and imp represent the contributions from the paramagnetic (PM) and impurity (imp) phases, respectively. Furthermore, $A_{\rm FM}$, $\lambda_{\rm FM}$, $f_{\rm FM}$, $\phi_{\rm FM}$ and $\lambda_{\rm FM}$ represent contributions from the internal FM field, together with $A_{\rm S}$, $\lambda_{\rm S}$. In particular, $A_{\rm FM}$ represents the internal magnetic field contributions that are perpendicular to the initial muon spin polarization, while $A_{\rm S}$ are contributions from the internal field that are parallel to the initial muon spin polarization. 

In order to properly separate the PM and the impurity phases under the applied TF, the constraint $\phi^{\rm PM}_{\rm TF}=\phi^{\rm imp}_{\rm TF}$ was set. Since an oscillating fraction of 0.08 was obtained at 110~K well below $T_{\rm C}$, the sample is found to contain about 34 $\%$ PM impurity phase, most likely Co$_2$P. Therefore, $A^{\rm imp}_{\rm TF}=0.0843(46)$ was fixed through the whole temperature range since this fraction should be temperature independent. Finally, the total asymmetry was fixed to 0.2351 for the measurements above 125~K, a value obtained from a high temperature measurement.

\begin{figure}[ht]
  \begin{center}
    \includegraphics[keepaspectratio=true,width=65 mm]{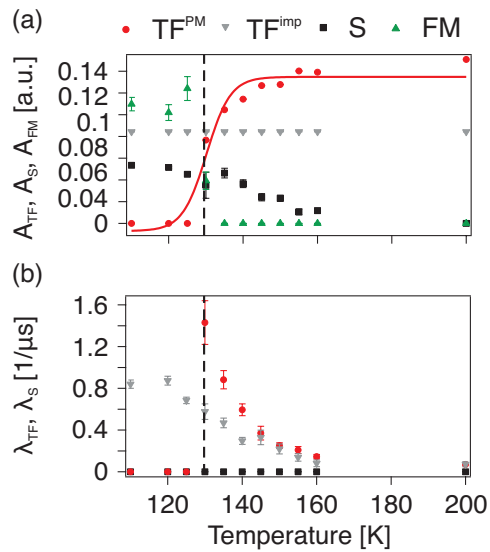}
  \end{center}
  \caption{Obtained fit parameters as a function of temperature using Eq.~(\ref{eq:TF}): 
  (a) Asymmetry and (b) transverse field and exponential relaxation rate. 
  The solid line in (a) is a fit using sigmoidal function. 
  The dotted vertical line across the figure indicates the transition temperature $T_{\rm C}=130.2(1.2)$~K.
  }
  \label{fig:La_TFPara}
\end{figure}

The obtained fit parameters using Eq.~(\ref{eq:TF}) with the procedure as described above are displayed in Fig.~\ref{fig:La_TFPara}. Each asymmetry component has a temperature dependence that is expected for a magnetically ordered sample. For $T<130$~K, $A^{\rm PM}_{\rm TF}=0$ and exhibits a sharp change at a certain critical temperature. Since $A^{\rm PM}_{\rm TF}$/($A_0-A^{\rm imp}_{\rm TF}$) roughly corresponds to the PM volume fraction, the abrupt change observed at $T\approx130$~K corresponds to the transition from a magnetically ordered state at low temperatures to magnetically disordered state at high temperatures. An accurate value of the transition temperature is obtained by fitting the $A^{\rm PM}_{\rm TF}(T)$ curve using a sigmoidal function, for which $T^{\rm TF}_{\rm C}=130.17(1.18)$~K is obtained. Similarly, $A_{\rm FM}=0$ above $T^{\rm TF}_{\rm C}$ while $A_{\rm S}$ poses none zero values even above $T^{\rm TF}_{\rm C}$, which steadily decreases with increasing temperature. Such behavior is naturally expected as these fractions stem from internal magnetic fields as described above. It should be noted that the $A_{\rm S}$ term is observed well above $T_{\rm C}$. The origin of such behavior is underlined in the ZF section presented below. 

The temperature dependencies of $\lambda^{\rm PM}_{\rm TF}$, $\lambda^{\rm imp}_{\rm TF}$ and $\lambda_{\rm S}$ are displayed in Fig. \ref{fig:La_TFPara}(b). $\lambda^{\rm imp}_{\rm TF}$ shows a steady decrease from lowest measured temperature up to the highest, and shows no anomaly at the magnetic transition. Such behaviour originate most likely from fluctuating Co $d$-moments and static Co and/or P nuclear moments, as also underlined in the ZF section. Such temperature dependence is similar to the one obtained in ZF configuration, underlying the quality of the fits in both field configurations. On the other hand, $\lambda^{\rm PM}_{\rm TF}$ is none zero only above $T_{\rm C}$, as expected. It has a maximum just above $T_{\rm C}$ and starts to decrease with increasing temperature, reflecting an increase in the internal field dynamics. Such temperature dependence will most likely follow the temperature dependence of magnetic susceptibility of a Curie-Weiss paramagnet. The small value of $\lambda_{\rm TF}$ (below $0.1~\mu{\rm s}^{-1}$) at high temperatures suggests that the PM fluctuations eventually become motionally narrowed for the $\mu^+$SR time window. 

The fact that $\lambda_{\rm S}=0~\mu$s$^{-1}$ across the whole measured temperature range suggests a static behavior of the internal FM field. Both $f_{\rm FM}$ and $\lambda_{\rm FM}$ represent the nature of the internal FM field but is not presented here. Instead, measurements in ZF configuration provide more accurate information regarding the internal fields. Finally, it should be noted that since $f_{\rm FM}\sim4~$MHz at 110 K, which is not very different from the TF precession frequency with 50~Oe (about 0.7~MHz), the estimated volume fraction of the nonmagnetic impurity phase includes an ambiguity. A more accurate value of the size of this fraction will be estimated using the ZF-$\mu^+$SR data.

\subsection{\label{sec:results}Zero field}
\begin{figure}[ht]
  \begin{center}
    \includegraphics[keepaspectratio=true,width=65 mm]{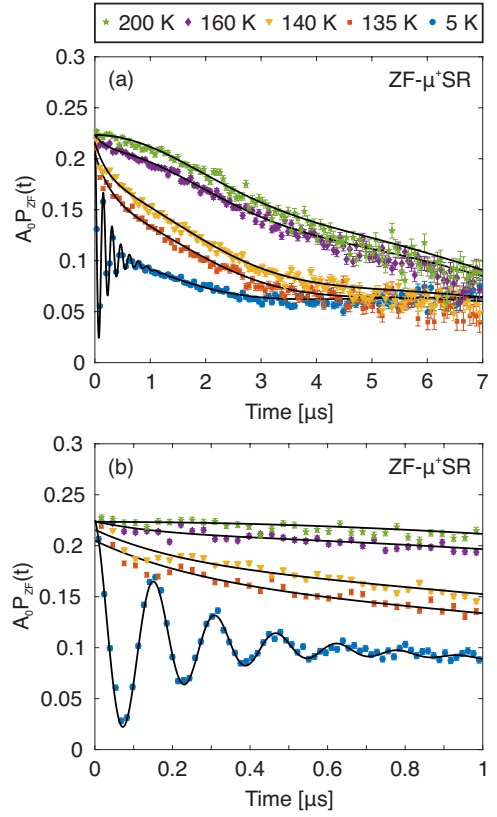}
  \end{center}
  \caption{ Zero field $\mu^+$SR spectra for selected temperatures 
  ($T=5$, 135, 140, 160 and 200~K) for the LaCaCo$_{2}$P$_{2}$ compound up to (a) $7~\mu$s and (b) $1~\mu$s. Solid lines represent the best fit using Eq.~(\ref{eq:ZF}).
  }
  \label{fig:La_ZFSpec}
\end{figure}

ZF-$\mu^+$SR time spectra for selected temperatures are shown in Fig.~\ref{fig:La_ZFSpec}. 
At 200~K, the ZF-spectrum in an early time domain exhibits a convex shape time dependence, which indicates a Gaussian-type relaxation.
A notable small dip around $3~\mu$s would suggest the presence of two independent Gaussian relaxations, consistent with the presence of an impurity phase. 
As temperature is lowered, the Gaussian relaxation is gradually changing into a more exponential like form, until an oscillation appears at lower temperatures below $T_{\rm C}$. 
In order to take into account all the phases present over the whole measured temperature range, the time spectra were fitted using a combination of two static Gaussian Kubo-Toyabe (SGKT) functions, two exponential relaxations, and one exponentially relaxing cosine oscillating terms:

\begin{eqnarray}
 A_0 \, P_{\rm ZF}(t) &=&
A_{\rm FM} \cos(f_{\rm FM}2\pi t+\phi_{\rm FM})e^{-\lambda_{\rm FM}t}\cr
&+& A_{\rm tail}e^{-\lambda_{\rm tail}t} \cr 
&+& A_{\rm F} e^{-\lambda_{\rm F}t} \cr 
&+& A_{\rm KT}G(t,\Delta_{\rm KT})e^{-\lambda_{\rm KT}t} \cr 
&+&A_{\rm imp}G(t,\Delta_{\rm imp})e^{-\lambda_{\rm imp}t} .
\label{eq:ZF}
\end{eqnarray}
$A_{0}$ is the initial asymmetry determined by the instrument's detector geometry and $P_{\rm ZF}$ is the muon spin polarization function in ZF configuration. The first two terms of Eq.~(\ref{eq:ZF}), i.e., $A_{\rm FM}$ and $A_{\rm tail}$, represents the response of the sample at low temperatures, when it enters a magnetically long range ordered state below $T_{\rm C}$. 
The third and forth terms, i.e., $A_{\rm F}$ and $A_{\rm KT}$, represents the sample response above or close to $T_{\rm C}$. 
The last term, $A_{\rm imp}$, corresponds to the impurity phase present in the sample, a constant term throughout the whole temperature range. In detail, $A_{\rm FM}$, $f_{\rm FM}$, $\phi_{\rm FM}$ and $\lambda_{\rm FM}$ are the asymmetry, frequency, phase, and relaxation rate resulting from perpendicular (with respect to the initial muon spin polarisation) internal field components, while $A_{\rm tail}$ and $\lambda_{\rm tail}$ are the asymmetry and relaxation rate of the tail component that inevitably exist in powder measurement of a magnetically ordered sample. This contribution stems from the fact that on average, 1/3 of the internal magnetic fields are parallel with respect to the initial muon spin polarisation for a perfect powder. $A_{\rm F}$ and $\lambda_{\rm F}$ are the asymmetry and relaxation rate of a fast component that manifests the ZF-spectra around $T_{\rm C}$. 
$A_{\rm KT}$ and $\lambda_{\rm KT}$ correspond to the asymmetry and relaxation rate related to the static Gaussian KT, represented by $G(t, \Delta)$ where $\Delta$ is the internal field distribution width stemming from isotropically distributed magnetic moments. 
The same description holds also for the subscript $A_{\rm imp}$ signal, but corresponds to the impurity phase instead of the main phase. 

In order to separate the various contributions, some constraints were set for the fits using Eq.~(\ref{eq:ZF}). 
In particular, $A_{\rm imp}=0.0485(9)$ and $\Delta_{\rm imp}=0.45(2)$ were fixed across the whole temperature range, as these parameters can be expected to be temperature independent. 
Such values were estimated at the base temperature given that the magnetic contrast between the main and impurity phase is the biggest. 
Indeed, the value of $\Delta_{\rm imp}$ suggests the presence of Co and/or P elements in the impurity phase. 
The value of $A_{\rm imp}$ corresponds to a volume fraction of 22~\%, 
which is lower than fraction estimated from TF configuration, as expected. 
It is noted that measurements (not shown here) in a longitudinal field configuration i.e. field parallel with the initial muon spin polarization, at 200 K confirm that the internal fields are static, supporting the fit with two static G-KT functions and that the internal fields originate from nuclear magnetic moments: $\Delta_{\rm KT}= 0.1264(12)~\mu$s$^{-1}$, which corresponds to 1.484237(2)~Oe.

\begin{figure}[ht]
  \begin{center}
     \includegraphics[keepaspectratio=true,width=65 mm]{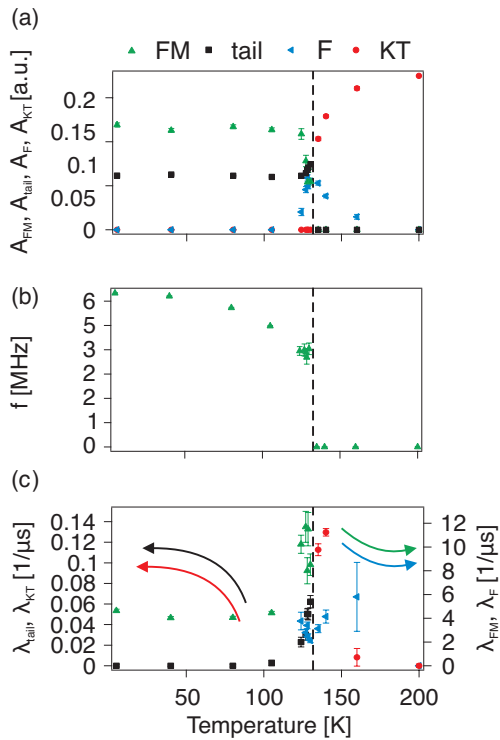}
  \end{center}
  \caption{Obtained fit parameters as a function of temperature using Eq.~(\ref{eq:ZF}): 
  (a) Asymmetry, 
  (b) precession frequency and 
  (c) relaxation rates. 
  The dotted vertical line across the figure indicates the transition temperature $T_{\rm C}=130.9(7)$~K.
  }
  \label{fig:La_ZFPara}
\end{figure}

The temperature dependencies of the obtained fit parameters are shown in Fig.~\ref{fig:La_ZFPara}, using the fit procedure as described above. 
Below $T^{\rm TF}_{\rm C}$, only two asymmetry components have none zero values (except for the $A_{\rm imp}$ that is not shown). 
This is consistent with the whole main sample phase entering a long range magnetically ordered state, for which $A_{\rm tail}\simeq\frac{1}{2}A_{\rm FM}$. 
Just below $T_{\rm C}$, a small upturn is observed in $A_{\rm tail}$ while a downturn is seen in $A_{\rm FM}$. 
This is natural given that parallel fluctuations usually increases as the static perpendicular component loses its structure close to $T_{\rm C}$. 
Intriguingly, the decrease of $A_{\rm FM}$ is followed by an additional fast relaxing component,  $A_{\rm F}$. 
Such component exhibits a maximum just below $T_{\rm C}$, for which $A_{\rm FM}$ is still none zero, and slowly decreases with further increasing temperature. 
Above $T_{\rm C}$, defined at the point where $f_{\rm FM}=0$~MHz and thus $A_{\rm FM}=0$, the spectra consist of $A_{\rm KT}$ and $A_{\rm F}$ components, where only $A_{\rm KT}$ persist above 160~K. 
It is noted that the fast exponential relaxation component ($A_{\rm F}$) is present between the temperature range 124 and 160~K, $i.e.$ across $T_{\rm C}$. 
This implies that long range and short range order coexist for a narrow temperature range below $T_{\rm C}$, 
which will be further discussed in Sec.~\ref{sec:discussion}.

The muon spin precession frequency ($f_{\rm FM}$), on the other hand, exhibits an order parameter like temperature dependence. 
The value of the frequency corresponds to the total magnitude of the local field at the muon site. 
A frequency of 6.33(2)~MHz is obtained at the base temperature and corresponds to 0.04670(15)~T, which can be considered relatively low. 
Detailed calculations referencing these values are presented in Sec.~\ref{sec:Mstructure}.


As for the relaxation rates, both $\lambda_{\rm FM}$ and $\lambda_{\rm tail}$ are temperature independent at low temperatures. 
Here, $\lambda_{\rm tail}$ corresponds to the spin-lattice relaxation rate. The value of $\lambda_{\rm tail}(2~$K$)=0~\mu$s$^{-1}$ suggests that the magnetic order at low temperatures is static. $\lambda_{\rm tail}$ increases as $T_{\rm C}$ is approached, indicating that 
the internal field is dynamic close to $T_{\rm C}$. 
$\lambda_{\rm FM}$ on the other hand corresponds roughly to the spin-spin relaxation rate and 
its value can be interpreted as the field distribution width at the muon site. 
Similar to the $\lambda_{\rm tail}$, an increase of $\lambda_{\rm FM}$ is observed close to $T_{\rm C}$. 
Part of it can be ascribed to increase in dynamics 
(in principle, $\lambda_{\rm FM}$ is composed of both spin-spin and spin-lattice relaxation rates).

The temperature dependencies of $\lambda_{\rm KT}$ and $\lambda_{\rm imp}$ are consistent with the findings in TF configuration. 
In Eq.~(\ref{eq:ZF}), each static Gaussian KT is multiplied by an exponential relaxation function. 
This is because the internal field at the muon site has two independent contributions (nuclear and electronic) and the time dependence of the muon polarization, $P(t)$, is given by the product of the expected polarization function originating from each contribution. 
Since the KT is the depolarization due to isotropically distributed nuclear moments, 
the exponential is accounting additional PM fluctuations present in the compound and it should follow the temperature dependence of $\lambda_{\rm TF}$.
Details of $\lambda_{\rm imp}$ are highlighted in Appendix~\ref{Appendix}. 
Finally, $\lambda_{\rm F}$ is the additional fast relaxation rate that manifests in the time spectra close to $T_{\rm C}$. 
Its temperature dependence shows a minimum close to $T_{\rm C}$, and increases again both below and above $T_{\rm C}$. 
The origin of this additional fast relaxation is discussed in Sec.~\ref{sec:discussion}.

\subsection{\label{sec:Mstructure}Magnetic structure}
For the sake of completeness, we have evaluated the internal field based on the FM structure proposed in Ref.~\onlinecite{Reehuis1994}. 
The magnetic structure was inferred from a combination of single crystal magnetization and polycrystal neutron diffraction (ND) measurements. 
In detail, ND measurements determined the $\mu_{\rm ord}=0.44(3)~\mu_{\rm B}$ to be aligned perpendicular to [001], 
which naturally cannot be resolved for tetragonal structures for a powder sample. 
Complementary magnetisation measurements determined then that the easy axis of magnetization to be along [100]. 

The expected muon sites for the given electrostatic potential and the corresponding local spin density (based on the FM structure as described above) was predicted within the Wien2K framework \cite{WIEN2k}. 
These calculations yield three possible moun sites; $\mu_1=(0,0,0.198)$, $\mu_2=(0,0.5,0.102)$ and $\mu_3=(0,0,0.5)$. 
The complete magnetic structure together with the predicted muon sites are shown in Fig. \ref{fig:La_Crystal} and 
the calculated local spin density at these sites are listed in Table~\ref{table:Fre}.

With this in mind, we attempt to calculate the local field at the predicted muon sites, based on the magnetic structure presented above. 
For an non-magnetized FM in ZF, the internal field at the moun site ($\bm B_{loc}$) is given by the following three components,

\begin{eqnarray}
\bm B_{loc} = \bm B_{\rm dip'} + \bm B_{\rm L} + \bm B_{\rm hf}
\label{eq:FM}
\end{eqnarray}
where $\bm B_{\rm dip'}$ is the resulting dipole field within the considered Lorentz sphere, 
$\bm B_{\rm L}$ is the Lorentz field, and 
$\bm H_{\rm hf}$ is the hyperfine contact field. 
Such internal fields can be translated into the corresponding precession frequency via $f=(\gamma_{\mu}/2\pi)|\bm B|$, so that 

\begin{eqnarray}
f_{loc} =\frac{\gamma_{\mu}}{2\pi} | \bm B_{\rm dip'} + \bm B_{\rm L} + \bm B_{\rm hf} |
\label{eq:fre_MF}
\end{eqnarray}
where $f_{loc}$ is the muon spin Larmor precession frequency around the internal field at the considered muon site. 
The modules of Eq.~\ref{eq:fre_MF} corresponds to the fact that $\mu^+$SR only detects the total magnitude of the local internal field.

In general, the local field components presented in Eq.~\ref{eq:FM} can be considered as contributions from localized and delocalized electrons. The hyperfine contact field accounts for delocalized electrons present at the muon site: $i.e.$ the local spin density at the muon site. 
The microscopic form of such interactions requires the detailed wave function of the electron. 
It is however common to assume that $\bm B_{\rm hf}$ is isotropic, $i.e.$ assuming a spherical electron wave functions. 
In this case, the $\bm B_{\rm hf}$ is simplified to

\begin{equation}
\bm B_{\rm hf}=\frac{2\mu_0}{3}| \psi(\bm r_\mu) | ^2 \bm m_e=\frac{2\mu_0}{3} \frac{\rho(\bm r_{\mu})}{| \bm m_e | }\bm m_e, 
\label{eq:Hyp}
\end{equation}
where $\mu_0$ is the vacuum permeability ($=4\pi\times$ $10^{-7}$~H/m) and 
the probability density for a spherical cloud at the muon site is given by $| \psi(\bm r_\mu) | ^2$, 
which in turn is related to the local spin density $\rho(\bm r_{\mu})$. 
In other words, $\bm B_{\rm hf}$ is given as a scalar coupling 
between $\rho(\bm r_{\mu})$ and the magnetic moment of the electron $\bm m_{e}=g\mu_{\rm B}\bm J$ 
where $\bm J$ is the total angular momentum.

The dipole field at the muon site on the other hand is originating from the dipolar interactions between localized electrons and the muon spin. 
A good approximation is to simply consider classical dipoles originating from spin polarized electron orbitals at the center of the magnetic atoms within a large sphere (the Lorentz sphere) with $N$ atoms:

\begin{eqnarray}
\bm B_{\rm dip'}= \frac{\mu_0}{4\pi}\sum^{N}_{j}\frac{3\bm r_{\mu j}(\bm m_{e,j}\cdot\bm r_{\mu j})}{r_{\mu j}^5}-\frac{\bm m_{e,j}}{r_{\mu j}^3},
\label{eq:Dip}
\end{eqnarray}
where if $N \to \infty$ then $\bm B_{\rm dip '}=\bm B_{\rm dip}$. $\bm r_{\mu,j}$ is the distance between the muon and the $j$-th ion. 
Since the summation in Eq.~(\ref{eq:Dip}) is not infinite but is instead limited up to within the so called Lorentz sphere, 
an additional contribution is added to the local field, known as the Lorentz field

\begin{eqnarray}
\bm B_{\rm L}=\frac{\mu_0}{3}\bm M_{\rm L}=\frac{\mu_0}{3V}\sum^N_j \bm m_{e,j}
\label{eq:L}
\end{eqnarray}
where $M_{\rm L}$ is the vector sum of the magnetic moments inside the Lorentz sphere divided by its volume. 

Based on presented theoretical models and the determined magnetic structure, 
the local fields were computed for the muon sites; (0,0,0.198), (0,0.5,0.102) and (0,0,0.5). 
The calculations were performed using Python package $MUESR$ \cite{Bonfa2017}, 
and the obtained local field values are presented in Table~\ref{table:Fre}. 
Among the considered muon sites, $\mu_2$ agrees well with the experimentally obtained data with $f_{\rm loc}=8.61\simeq6.332(23)=f_{\rm FM}(5$~K$)$,  although the calculated value slightly overestimates the local field in comparison to the data. 
This is most likely due to local magnetic excitations present in a FM, 
according to the Bloch-3/2 law~\cite{Blundell2003, Forslund2020_La}, 
effectively lowers the precession frequency due to the spontaneous magnetization. 

The Lorentz field is independent on muon site, 
since the considered Lorentz sphere was kept constant for the all muon sites. 
It is however noted that given the symmetry of the positions of $\mu_2$ and $\mu_3$, 
$\bm B_{\rm dip'}$ and $\bm B_{\rm L}$ are effectively canceling each other 
resulting into a low precession frequencies, especially for $\mu_3$. The hyperfine contact fields are fairly constant across the considered muon sites and constitute a fairly large portion of the resultant local field. 
Such behavior is different from the $A$-type AF NaNiO$_2$~\cite{Forslund2020}, 
where the local field was found to be solely formed by dipolar fields. This is to some degree expected since the local spin density should be more considerable in a FM, compared to an AF.

\begin{table*}[ht]
\caption{\label{table:Fre}
The calculated local field values at the predicted muon sites are tabulated together with the obtained local spin density $\rho(r_{\mu})$ for the given magnetic structure. 
$f_{\rm loc}$ was evaluated using Eq.~(\ref{eq:fre_MF}), 
whereas the dipole field, Lorenz field and the hyperfine contact field are given by Eq.~(\ref{eq:Dip}), Eq.~(\ref{eq:L}) and Eq.~(\ref{eq:Hyp}), respectively. 
The experimentally obtained precession frequency $f_{\rm FM}(5$~K$)$ and the expected internal field distribution width are listed as well. 
}

\begin{ruledtabular}
\begin{tabular}{lccccccc}
 Muon site & $\rho(\bm r_{\mu})~[\mu_{\rm B}$\AA$^{-3}]$&$\bm B_{\rm dip'}$~[T]&$\bm B_{\rm hf}$~[T]&$\bm B_{\rm L}$~[T]&$f_{\rm loc}$~[MHz] &$f_{\rm FM}(5$~K$)$~[MHz] & $\Delta^{\rm calc}_{\rm LaP_2Co_2}~ [\mu{\rm s}^{-1}]$\\
 \hline
 $\mu1$~(0,0,0.198) & -0.00237581 &[0.0872,0,0]&[-0.0185,0,0]&[0.0429,0,0]&15.13&  6.332(23)& 0.379\\
 \hline
 $\mu2$~(0,0.5,0.102) &  -0.0015081
 &[-0.0947,0,0] &[-0.0117,0,0]&[0.0428,0,0] &8.61&   6.332(23) & 0.461\\
 \hline
 $\mu3$~(0,0,0.5) &-0.00176282  &[-0.0379,0,0]&[-0.0137,0,0]     &[0.0428,0,0]&1.86&   6.332(23) & 0.297\\
\end{tabular}
\end{ruledtabular}
\end{table*}

\section{\label{sec:discussion}Discussion}
\begin{figure}[ht]
  \begin{center}
     \includegraphics[keepaspectratio=true,width=79 mm]{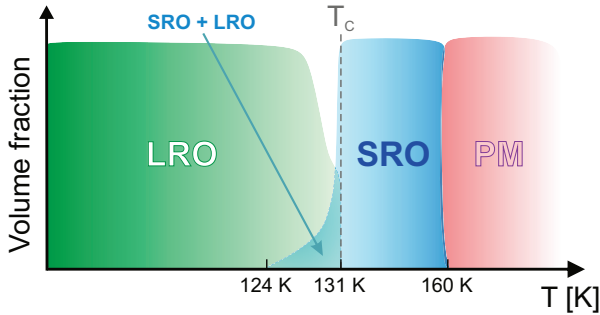}
  \end{center}
  \caption{Magnetic phase diagram as a function of temperature, with $T_{\rm C}=130.91(65)$. A long range magnetically ordered phase (LRO) for $T<124$~K, a short range magnetically ordered phase (SRO) for $T_{\rm C} \leq T \leq 160$~K and a paramagnetic phase (PM) for $T>160$~K are stabilised as a function of temperature. A temperature regions in which LRO and SRO co-exist present itself for $124\leq T \leq T_{\rm C}$. The magnetic volume fractions are estimated based on Fig.~\ref{fig:La_ZFPara}(a).
  }
  \label{fig:La_phase}
\end{figure}

The value of the local field at the $\mu_2$ site at low temperatures is reasonably explained by the magnetic structure  determined by neutron diffraction. Therefore, we will to focus our discussion on the behavior of the compound at higher temperatures. 
The ZF scan presented above reveals that the sample undergoes a cascade of magnetic transitions (Fig.~\ref{fig:La_phase}). 
In detail, a PM order is established at higher temperatures $T>160$~K, 
as evidenced by the exponentially relaxing KT ($A_{\rm KT}$). 
A short range order is stabilized in the temperature range $T_{\rm C}<T\leq160$~K, 
for which the muon spin depolarization is made up of two separate exponential relaxations 
($A_{\rm F}$ and $A_{\rm KT}$ with $\Delta=0$~$\mu s^{-1}$). 
Intriguingly, such short range order (SRO) is likely to coexist with the FM long range order (FM-LRO) 
in the range $124$~K$\leq T\leq T_{\rm C}$, 
for which FM-LRO is fully formed below $T<124$~K. 
Such assertion is supported by the fact that $A_{\rm F}$ is split into two components at $T_{\rm C}$. 
While $A_{\rm KT}$ settles to a value $\sim A_0/3$, consistent with the tail component, 
$A_{\rm F}$ shows a maximum at $T_{\rm C}$ which is separated into $A_{\rm FM}$ and $A_{\rm F}$ just below $T_{\rm C}$. 
Since $A_{\rm FM}$ is the oscillating fraction, this is the long range ordered fraction of the sample. 
$A_{\rm F}$ on the other hand is still an exponential relaxation even below $T_{\rm C}$, 
implying that there is a small temperature range in which FM-LRO and SRO co-exist. 

While it was not clearly commented on, 
a similar behavior was observed in $^{31}$P-NMR study in the raw data figure~\cite{Imai2015}. 
In fact, the spin-spin relaxation rate ($1/T_2$) seems to decrease between 130 and 120~K, 
the $1/T_2$ increases again around 110~K. 
Such behavior is reflected by the behavior of $\lambda_{\rm F}$, presented in Fig.~\ref{fig:La_ZFPara}(c). 
Although NMR does not provide the information on the volume fraction, 
our study clearly show that a small fraction of SRO is present while the majority of the sample forms LRO. 

A coexistence of LRO and SRO has been reported for SrEr$_2$O$_4$ \cite{Hayes2011} or La$_{0.7}$Sr$_{0.3}$Mn$_{1-x}$Co$_x$O$_3$ \cite{Thanh2015}. 
SrEr$_2$O$_4$ is a frustrated magnet where the coexistence was attributed to complex interactions 
arising from geometrically frustrated magnetism. 
For La$_{0.7}$Sr$_{0.3}$Mn$_{1-x}$Co$_x$O$_3$, on the other hand, 
the coexistence looks to stem from the competitions between AF and FM interactions. 
This would suggest that similar competitions among multiple interactions should be present in LaCo$_2$P$_2$ as well. 
In fact, LaCo$_2$P$_2$ shows a strong anisotropic character in the FM state, 
because of the 2D interactions in the CoP plane \cite{Imai2015}. 
At high temperatures, on the other hand, the fluctuation were shown to be of a 3D character \cite{Imai2015}. 
In other words, there exists a small temperature range for which there is a competition between the 2D and 3D character of the interactions, resulting into the coexistence of SRO and LRO. 
A similar seperation of ordered and disordered phases has also been reported for Cu$_2$IrO$_3$ \cite{Kenney2019} and Li$_2$RhO$_3$ \cite{Khuntia2017}. 
Theoretical treatment of such phenomena revealed that the ground state is stabilized from competition between quantum fluctuation and frustration \cite{Gonzalez2019, Seifert2019}. 

We wish to note the fact that a relatively large impurity phase is present in the compound, 
which could be the origin behind the anomaly around $T_{\rm C}$. 
However, the signal from the impurity phase is nicely fitted with a exponentially relaxing Gaussian KT at 2~K, 
and it is highly unlikely that such behaviour should evolve into an exponential like relaxation at higher temperatures. 
The value of the exponential relaxation rate itself is also one order of magnitude lower than the one of $\lambda_{\rm F}$ (Appendix~\ref{Appendix}). 
Moreover, $\mu^+$SR is sensitive to detecting magnetic volume fractions, 
meaning each contribution is in principle separated. 
We wish to stress that the XRD analysis indicated no structurally phase separations, $i.e.$ the sample is chemically homogeneous. The previous study underlines the high quality of our sample \cite{Imai2015}. 



\section{\label{sec:conclusion}Conclusions}
Transverse and zero field (ZF) $\mu^+$SR measurements reveal that the sample exhibits a long range ferromagnetic (FM) ground state below $T_{\rm C}=130.91(65)$~K, consistent with previous reports. 
The muon sites and the corresponding local spin density, based on the already reported magnetic structure, were predicted by density functional theory. 
The estimated local field calculations agrees well with the presented ZF-$\mu^+$SR data. 
Intriguingly, this study reveals cascade of magnetic transitions, not observed in previous studies. 
In detail, a paramagnetic (PM) phase is found at higher temperatures $T>160$~K. 
A short range order (SRO) is stabilized at lower temperatures, 
for which a coexistence of the SRO and long range FM is present for $124~$K$\leq T\leq T_{\rm C}$. 
The coexistence originates from a competition of 2D and 3D magnetic interactions and/or fluctuations, prominent at $T_{\rm C}$,  
since a coexistence of magnetic phases is commonly found in compounds with competing magnetic interactions. 

\begin{acknowledgments} 
We thank the staff of PSI for help with the $\mu^+$SR experiments. We also appreciate H. Nozaki, M. Harada and R.~Scheuermann for their help with the $\mu^+$SR experiments. This research was supported by the Swedish Research Council (VR) (Dnr. 2016-06955) as well as the Swedish Foundation for Strategic Research (SSF) within the Swedish national graduate school in neutron scattering (SwedNess). J.S. acknowledge support from the Ministry of Education, Culture, Sports, Science and Technology (MEXT) of Japan, KAKENHI Grant No.23108003 and Japan Society for the Promotion Science (JSPS) KAKENHI Grant No. JP26286084, JP18H01863 and JP20K21149. D.A. acknowledges partial financial support from the Romanian UEFISCDI project PN-III-P4-ID-PCCF2016-0112 (6/2018). H.O. acknowledge support from KAKENHI Grant No. JP20K05663. The crystal figure was drawn using VESTA \cite{Momma2008}.

\end{acknowledgments}

\appendix
\section{Temperature dependence of the impurity phase}
\label{Appendix}
\begin{figure}[ht]
  \begin{center}
    \includegraphics[keepaspectratio=true,width=65 mm]{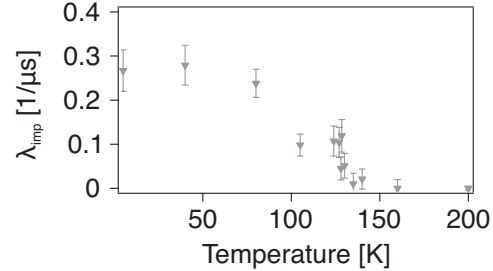}
  \end{center}
  \caption{Temperature dependence of the KT relaxation rate for the impurity phase
 }
  \label{fig:La_lamKT_imp}
\end{figure}

The temperature dependence of the KT relaxation rate for the impurity phase, $\lambda_{\rm imp}$, is shown in Fig.~\ref{fig:La_lamKT_imp}. The values saturates at lower temperatures and steadily decreases with increasing temperature. As mentioned in the main text, the temperature dependence follows the one obtained in TF configuration, underlying the high quality of our fits in both field configuration. Moreover, the value itself is one order of magnitude lower than that of $\lambda_{\rm F}$, suggesting that the origin of the additional exponential close to $T_{\rm C}$ is not from the impurity but is truly an intrinsic behavior of LaCo$_2$P$_2$.

\bibliography{Refs} 
\end{document}